\documentclass[sigconf]{acmart}

\usepackage{booktabs} 
\usepackage{subcaption}
\usepackage{fancyhdr}

\setcopyright{none}

\renewcommand\footnotetextcopyrightpermission[1]{} 
\makeatletter
\renewcommand\@formatdoi[1]{\ignorespaces}
\makeatother

\acmDOI{}

\acmISBN{}

\acmConference[]{}{}{} 
\acmYear{2018}
\copyrightyear{2018}

\begin{document}
\pagestyle{fancy}
\lhead{\textit{[UNDER SUBMISSION]}}
\title{On Using Non-Volatile Memory in Apache Lucene}

\author{Ramdoot Pydipaty}
\affiliation{%
    \institution{Cisco Systems, Bangalore, India}
}
\email{rapydipa@cisco.com}

\author{Amit Saha}
\affiliation{%
    \institution{Cisco Systems, Bangalore, India}
}
\email{amisaha@cisco.com}

\begin{abstract}
Apache Lucene is a widely popular information retrieval library
used to provide search functionality in an extremely wide
variety of applications. Naturally, it has to efficiently index and search 
large number of documents. 
With non-volatile memory in DIMM form factor (NVDIMM),
software now has access to durable,
byte-addressable memory with write latency within
an order of magnitude of DRAM write latency.

In this preliminary article,
we present the first reported work on the impact of using NVDIMM
on the performance of committing, searching, and near-real time searching
in Apache Lucene. We show modest improvements by using NVM but,
our empirical study suggests that bigger impact requires
redesigning Lucene to access NVM as byte-addressable
memory using {\it loads} and {\it stores},
instead of accessing NVM via the file system.
\end{abstract}


\maketitle

\section{Introduction}
\label{introduction}

Non-Volatile Memory (NVM)~\cite{acmq:nvm}, also called Persistent Memory (PMEM) 
or Storage Class Memory (SCM), is a disruptive trend in the compute technology
landscape. In addition to providing data durability (traditionally
provided by HDDs/SSDs), these devices also behave like memory (DRAM) by providing
byte addressability, thus giving applications the ability to access durability via
{\it load/store} operations. Moreover,
NVM provides access speeds closer
to that of DRAM\footnote{500ns for 3D-XPoint DIMM compared to 100ns 
for DRAM and 30microsec for SSD~\cite{latency:url}}
\footnote{Since 3D-XPoint in DIMM form factor is still not
available, the numbers for 3D-XPoint are speculative and based on
marketing documents.},
much faster than traditional
secondary storage technologies, including SSDs.


Apache Lucene~\cite{lucene:url} is very widely used,
open source,
Java based\footnote{ports and integrations to C/C++, C\#, Ruby, Perl, Python, etc.
are available.},
high-performance information retrieval library. Since it is 
not a complete application by itself, applications implement search
functionality (such as indexing, querying, and highlighting)
by using the APIs exposed by Lucene. Being a
text search library, Lucene has to efficiently
handle a large number of documents
being written and indexed (and subsequently searched).
As we describe later in Section~\ref{lucene} 
that in order for faster performance, Lucene does not necessarily
commit the data to durable storage, thus sacrificing persistence of data to some extent.
This, coupled with
the fact that Lucene is the search library used by the 
highly popular search engines
(e.g., Elasticsearch~\cite{elasticsearch:url} and Apache Solr~\cite{solr:url}),
it is a natural to ask,
{\it "What is the impact of using non-volatile memory on Apache Lucene?"}

To the best of our knowledge, this is the first work that explores
the use of NVM in Apache Lucene.
There is some prior work on the effect of persistent memory on
distributed storage systems. Islam et al.~\cite{islam:ics16} used
NVM with RDMA in
HDFS~\cite{hdfs} to utilize the byte-addressability of NVM. There has been
some work~\cite{debrabant14, arulraj17, mnemonic:url}
on designing database systems to take advantage of NVM.
However, these are not directly relevant to the work presented
in this paper.

Our contributions in this paper are as follows:
\begin{itemize}
\itemsep0em 
\item We present the first reported study of using NVM in Apache Lucene.
\item We quantify the impact of using NVM on indexing, searching, and near-real-time
searching in Apache Lucene.
\item We identify fundamental changes needed in the operational model of
Apache Lucene to maximize the impact of NVM on Lucene's performance.
\end{itemize}

The rest of the paper is organized as follows. We provide an overview of Apache Lucene
in Section~\ref{lucene}. Section~\ref{evaluation} presents the details of the evaluation
and we draw conclusions in Section~\ref{conclusions}.
\section{Lucene Overview}
\label{lucene}
\begin{figure}
    \centering
    \includegraphics[width=0.8\linewidth]{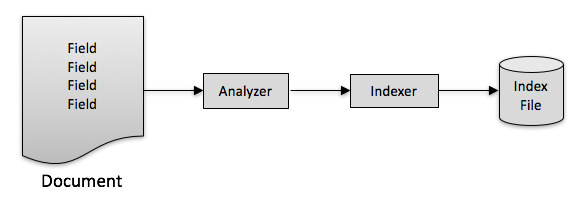}
    \caption{Indexing steps in Lucene~\cite{lucene_in_action:book}.}
    \label{fig:indexing}
\end{figure}

In this section we give a brief overview of Lucene; the interested reader
can refer to citations~\cite{lucene4,lucene_in_action:book} for more detailed
descriptions. Here we focus mostly on the parts that are relevant to this paper.

\subsection{Indexing}
\label{indexing}
Figure~\ref{fig:indexing} shows the typical processing done by Lucene
to documents fed to it by the encompassing application;
Lucene itself cannot acquire content.
During indexing, the text is extracted from the original content and a \texttt{Document}
is created that contains \texttt{Fields} holding the content. The contents of the \texttt{Fields}
is passed through the \texttt{Analyzer}
to generate a stream of tokens, which are passed through the \texttt{Indexer} to generate an
{\it inverted index}.
Finally, the index is written to either a new index file
or the information is added/removed/updated to an existing index file. These index files
are stored in a user specified {\it directory}.

In order to be fast, index files are {\it immutable}, thus having no requirement of locking and hence
avoiding any costly synchronization between multiple writer threads.
However, immutability
usually means that in order to update an index (say a particular document no longer has a term it
initially contained), an entirely new index would have to be created and the old one deleted.
In reality, in order to be efficient, 
a Lucene index is made up of multiple {\it immutable} index segments,
On a search, Lucene searches over all segments for an index,
filters out any deletions\footnote{since the segments are immutable, a document might have been
initially added to  and index and then deleted before the search is performed},
and finally, merges the results from all the segments. As explained in the next section, the
immutability of segments has implications on search.

\begin{figure}[htp]
     \begin{subfigure}[b]{0.495\linewidth}
        \includegraphics[width=\linewidth]{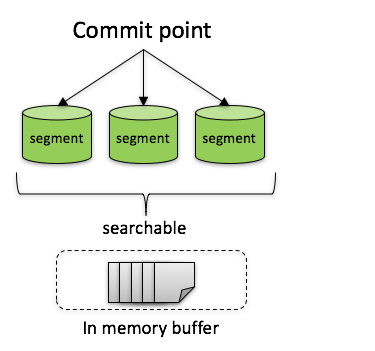}
        \caption{Before flushing.}
        \label{fig:before_flush}
    \end{subfigure}
    \begin{subfigure}[b]{0.495\linewidth}
        \includegraphics[width=\linewidth]{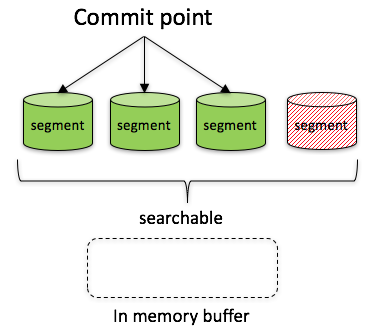}
        \caption{After flushing in NRT.}
        \label{fig:after_flush}
    \end{subfigure}
    \caption{Index updates in Lucene. Even after a flush, data is not committed
    to durable storage.}
\label{fig:segment_states}
\end{figure}

\subsection{Searching}
\label{searching}
The basic search process is quite straightforward. 
Once indexing of a document is {\it complete}, searching takes in a
{\it directory} (Section~\ref{indexing})
and a {\it term = <fname, fvalue>}, which is the basic unit for searching,
and searches for documents that contain ``fvalue'' in a field with name ``fname''.

However, there are some nuances of search due to the fact that index segments are immutable.
Figure~\ref{fig:before_flush} shows a scenario in which
the \texttt{Commit point} refers to a state of the index segments that are 
{\it committed} in disk and hence are durable. However, in a live system, additional documents are 
constantly being analyzed and indexed, thus necessitating the creation of new index segments. 
The indexing information for new documents are first written in an {\it in-memory buffer} and then 
later {\it committed} to the underlying persistent storage, at which point an updated 
\texttt{Commit point} is generated.
As can be imagined, this process is expensive due to the requirement of
{\it fsync'ing} the data to disk.

Consequently, till
the new segment is not stored in disk, the segment {\it first}, cannot be searched since uncommitted 
data cannot be part of the \texttt{Commit point} and {\it second}, the indexing buffer is volatile
since it is simply stored in DRAM. This makes the in-memory data (in a single Lucene instance)
susceptible to system or power failure\footnote{Search applications such as ElasticSearch need to 
implement redundancy so that failure of a single Lucene node cannot make the data unavailable},
in addition to adding delay between processing a document and being able to search the document.
This delay, though small (in the order of minutes), is not acceptable to many
applications and led to the development of the Near Real Time (NRT) search as in
Section~\ref{nrt}.

\subsection{Near Real Time (NRT) Search}
\label{nrt}
As shown in Figure~\ref{fig:after_flush}, NRT search is achieved by ensuring that the new index
segment is not written directly to the disk but instead to the {\it filesystem cache} -- and only
later committed to disk. Once the segment is written to the filesystem, for all practical
purposes, the segment is made searchable.
This allows newer files to be indexed and searched without requiring a 
{\it full commit}. This reduces the time between indexing a document and being able to search
the document, {\it though at the risk of losing the data in the event of a system failure}.
The application can force the flushing of the in-memory buffer (so that the data becomes searchable)
by calling the {\it reopen} API.

\section{Evaluation}
\label{evaluation}
In this section we describe the experiments that we performed to
quantify the impact of using NVM in Lucene. We start with describing
our experimental setup.

\subsection{Experimental Setup}
\label{setup}
We used a single machine with 2.6~GHz Intel Xeon CPU,
with 28~cores, 56~vCPUs (hyper threading enabled),
1~TB DDR4 RAM (running at 2.4~GHz), a 2~TB SSD that is accessed
over a SATA3.0 (6~Gbps) interface.
Since NVDIMMs are not yet commercially available,
we carved out a space of 768~GB from the 1~TB RAM 
for a pmem device at \texttt{/dev/pmem}
using the kernel's {\it mmap}
settings as explained in the Persistent Memory Wiki~\cite{pmem:url}
\footnote{This is standard practice in doing preliminary evaluation
using NVDIMM.}.
On the pmem device we placed an ext4 filesystem with DAX extension.
We could not use {\it mmap} (load/store) for our experiments
as {\it Lucene uses a file abstraction and not a device abstraction}.
Instead we used the filesystem abstraction, i.e., the pmem device was accessed
via the kernel's file system code.
We used the trunk version of Lucene (i.e., $>7.2.1$)
and used the latest version of the {\it luceneutil} benchmark in our
experiments.
\subsubsection*{Luceneutil Benchmarking Tool}
The {\it luceneutil}~\cite{luceneutil:url} tool is 
the de-facto benchmark utility for Apache Lucene.
It indexes the entire Wikipedia English export~\cite{wikipedia_data}
file as the input dataset. It runs benchmarking tests for Indexing,
Searching , NRT, Sorting, Geobench and a  variety of other tests\footnote{
The results of the {\it luceneutil} benchmark run on the Apache Lucene trunk is reported
on a nightly basis at https://home.apache.org/~mikemccand/lucenebench/index.html.}.
To investigate the effects of pmem on lucene,
we chose to run a subset of benchmark tests from luceneutil.
The set of benchmark tests executed were indexing performance benchmark,
search performance benchmark, and NRT performance benchmark.
 
For regular case, index files are stored on ext4 file system backed by SATA3.0 SSD.
While for PMEM case, index files are stored on ext4 file system backed by PMEM device

\subsection{Indexing Performance}
\label{eval:indexing}
\begin{figure}[!tbp]
    \centering
    \includegraphics[width=0.6\linewidth]{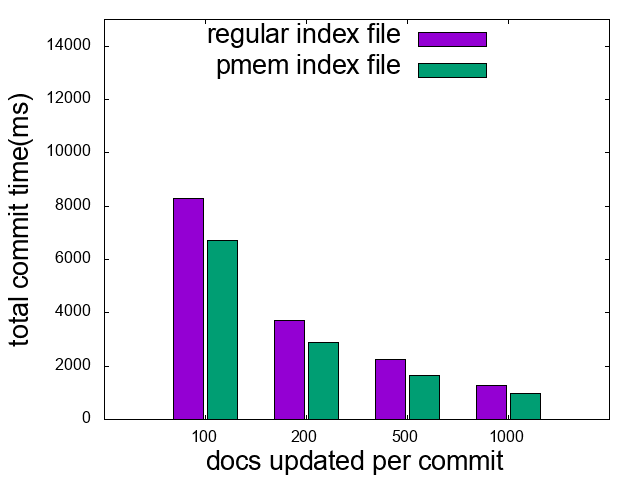}
    \caption{Commit performance.}
    \label{fig:commit}
\end{figure}
\begin{figure}[!htbp]
     \begin{subfigure}[b]{0.495\linewidth}
        \includegraphics[width=\linewidth]{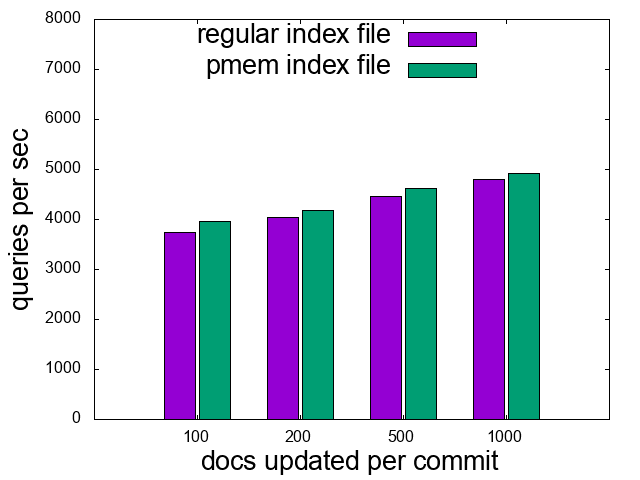}
        \caption{Queries per second.}
        \label{fig:qps}
    \end{subfigure}
    \begin{subfigure}[b]{0.495\linewidth}
        \includegraphics[width=\linewidth]{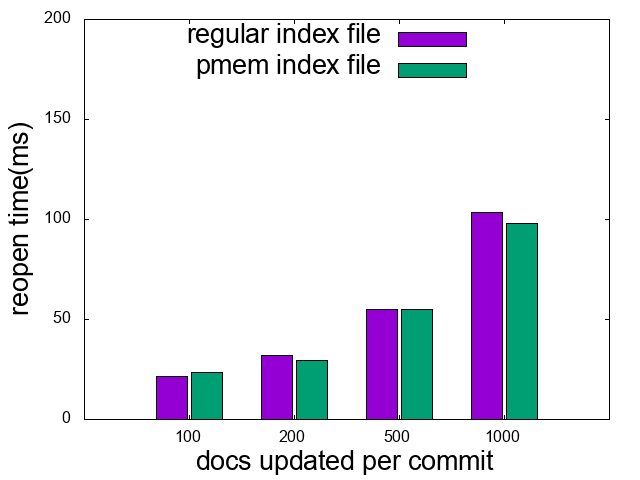}
        \caption{Reopen time.}
        \label{fig:reopen}
    \end{subfigure}
    \caption{Queries per second and Reopen times for NRT benchmark.
    Regular index file is stored in SSD.}
\label{fig:nrt}
\end{figure}

\begin{figure*}[!tbp]
    \centering
    \includegraphics[width=\linewidth]{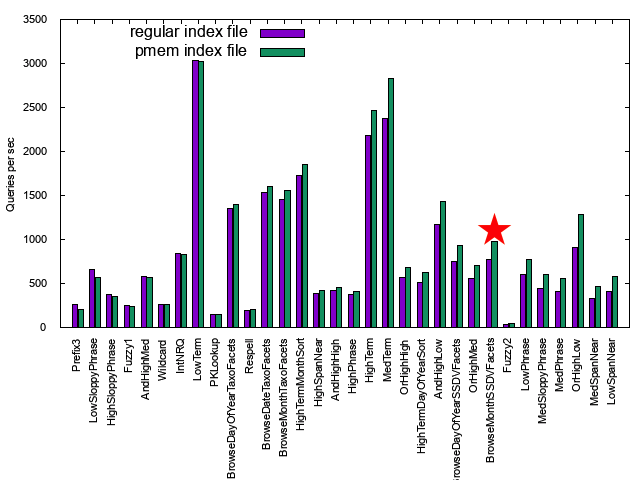}
    \caption{Search Bench Performance, arranged in increasing gains from left to right.}
    \label{fig:search}
\end{figure*}

This benchmark indexes the wikimedium500k\footnote{A collection of 500k lines from
the main source that has millions of lines.} from the
Wikipedia English export~\cite{wikipedia_data}.
We mapped the indexed file to regular (SSD) as well pmem backed files.
As the indexing part itself is compute intensive and is not dependent on the
underlying storage technology, we separated the total time for indexing as
index computation time and commit time.
In both the cases, the index computation times are same (we do not show the results here).
As shown in Figure~\ref{fig:commit}, with NVM,
the commit times show improvement in the range of 20\%--30\%
across all frequencies of commits (number of docs/commit). The improvement is more
pronounced with higher frequency of commits (100~docs updated/commit) since the amount
of data to be written is smaller. With larger writes, even SSD shows good performance
since entire, large segments can be written sequentially, thus eliminating much of the advantage
of random and/or small writes to NVM.

\subsection{Search Performance}
\label{eval:bench}

Search performance benchmark of luceneutil covers families of search tests ---
BooleanQuery, ProximityQueries, FuzzyQuery, Faceting, Sorting, etc. 
We ran these search benchmark tests against indices stored in regular files and 
pmem files. As shown in Figure~\ref{fig:search},
we observed that for PMEM use case, 12 tests (out of 32 tests in the bench) 
achieved considerable gains (>20\%) in the queries handles per second,
while another 12 saw gains of
$\leq$20\%, and the remaining didn't see any gain (or saw marginal loss in performance). 

We noticed that some of the tests achieved significant gains, $\geq$25\%. 
For instance, ``BrowseMonthSSDVFacets'' (marked with a star) is one of them. 
On closer inspection we found that this test executes
a query over the summarized/aggregated data of month field. 
It essentially covers the {\it Doc Values (DV)} feature of lucene. 

Inverted index is good at finding documents that contain a term. 
It does not perform well in the opposite direction --- 
determining which terms exist in a single document. 
Doc Values are a representation of documents (uninverted index), 
which aid in handling these queries. They are generated at index-time and 
serialized to disk. It relies on the OS file system cache to manage memory 
instead of retaining structures on the JVM heap. 
In situations where the ``working set'' of data is smaller than the available memory, 
the OS will naturally keep the doc values resident in memory. 
But when the data is much larger than available memory, 
the OS will begin paging the doc values on/off disk as required, 
making it much slower.  
In these cases, we can expect PMEM to deliver significant 
gains compared to regular disk.


\subsection{NRT Search Performance}
\label{eval:nrt}

Finally, Figure~\ref{fig:nrt} shows the results of our experiments with NRT using the
NRT performance benchmark, included in the luceneutil tool.
The test is run with one thread each for indexing, search, and reopen. As explained in
Section~\ref{nrt}, the {\it reopen} API call forces the in-memory buffer to be emptied and the
contents made searchable though not committed to disk yet.
The documents are updated at a constant rate of 1000~docs/sec whereas one {\it reopen()}
request is done every second. The entire test is run for 60~seconds and we vary the 
commit (the step that actually makes the data durable by writing to either SSD or PMEM) frequency
from 100~docs/commit (frequent commits) to 1000~docs/commit (infrequent commits). We measure
how many queries/second can be handled and how long it takes to perform
the {\it reopen} call.

Figure~\ref{fig:qps} shows that the number of queries per second that the system can support
increases as the frequency of commits decreases. 
This can be explained by the fact that every time a commit is done,
the in-memory buffer gets flushed
to a segment and the segment is written (fsync'ed) to the underlying storage. This then leads to the 
creation of a new \texttt{Commit point}. Thus doing commits frequently hurts the ability to respond
to queries quickly.

Figure~\ref{fig:reopen} shows the time it takes to complete a reopen call, i.e.,
to flush the in-memory buffer to a segment in the file system cache (but not yet committed to storage).
Here frequent commits helps because the in-memory buffer gets cleared out more frequently when the 
commit frequency is higher (100~docs/commit) and hence a reopen call has to copy less information
from the in-memory buffer to form a segment in the file system cache. This allows the reopen call to
complete faster, thus improving the reopen time with more frequent commits.

However, the most important thing to note in both the NRT results is that there is 
{\it negligible performance difference between storing the index file in SSD vs. a pmem device}.
In hindsight, the reason is clear --- since Lucene uses a file system abstraction, the file system cache
is able to service most of the search requests, thus masking the difference in speed between the pmem
and the SSD device. Though committing frequently obviously would make the data durable faster, 
these benchmark tests cannot expose the benefit of durability. 
{\it Future research in this area would have
to figure out reliable benchmarks to quantify the impact of durability (or the lack thereof)}.


\section{Conclusions and Future Work}
\label{conclusions}
We presented a preliminary, and, to our knowledge,
{\it the first} work that explores the impact of 
non-volatile memory in Apache Lucene, a very popular
information retrieval library.

Our experiments show modest gains by pointing Lucene
index segment files to the NVM device. However, this is just the tip
of the iceberg and higher gains may be achieved by say, implementing
index segments that bypasses the file system entirely and instead
is read/written directly into NVM using loads/stores.
{\it We believe that future work in this direction will lead to
redesigning at least parts of Lucene (and other similar libraries)
to bypass the file system
and directly access non-volatile memory using loads and stores}.

\bibliographystyle{ACM-Reference-Format}
\bibliography{ms} 


\begin{thebibliography}{15}


\ifx \showCODEN    \undefined \def \showCODEN     #1{\unskip}     \fi
\ifx \showDOI      \undefined \def \showDOI       #1{#1}\fi
\ifx \showISBNx    \undefined \def \showISBNx     #1{\unskip}     \fi
\ifx \showISBNxiii \undefined \def \showISBNxiii  #1{\unskip}     \fi
\ifx \showISSN     \undefined \def \showISSN      #1{\unskip}     \fi
\ifx \showLCCN     \undefined \def \showLCCN      #1{\unskip}     \fi
\ifx \shownote     \undefined \def \shownote      #1{#1}          \fi
\ifx \showarticletitle \undefined \def \showarticletitle #1{#1}   \fi
\ifx \showURL      \undefined \def \showURL       {\relax}        \fi
\providecommand\bibfield[2]{#2}
\providecommand\bibinfo[2]{#2}
\providecommand\natexlab[1]{#1}
\providecommand\showeprint[2][]{arXiv:#2}

\bibitem[\protect\citeauthoryear{??}{hdf}{[n. d.]}]%
        {hdfs}
 \bibinfo{year}{[n. d.]}\natexlab{}.
\newblock \bibinfo{title}{Apache {Hadoop HDFS}}.
\newblock \bibinfo{howpublished}{\url{https://hortonworks.com/apache/hdfs/}}.
  (\bibinfo{year}{[n. d.]}).
\newblock


\bibitem[\protect\citeauthoryear{??}{luc}{[n. d.]a}]%
        {lucene:url}
 \bibinfo{year}{[n. d.]}\natexlab{a}.
\newblock \bibinfo{title}{Apache Lucene Core}.
\newblock \bibinfo{howpublished}{\url{https://lucene.apache.org/core/}}.
  (\bibinfo{year}{[n. d.]}).
\newblock


\bibitem[\protect\citeauthoryear{??}{mne}{[n. d.]}]%
        {mnemonic:url}
 \bibinfo{year}{[n. d.]}\natexlab{}.
\newblock \bibinfo{title}{Apache {Mnemonic}: An Open Source Java-based
  storage-class memory oriented durable object platform for linked objects
  processing and analytics}.
\newblock \bibinfo{howpublished}{\url{https://mnemonic.apache.org/}}.
  (\bibinfo{year}{[n. d.]}).
\newblock


\bibitem[\protect\citeauthoryear{??}{sol}{[n. d.]}]%
        {solr:url}
 \bibinfo{year}{[n. d.]}\natexlab{}.
\newblock \bibinfo{title}{Apache Solr}.
\newblock \bibinfo{howpublished}{\url{http://lucene.apache.org/solr/}}.
  (\bibinfo{year}{[n. d.]}).
\newblock


\bibitem[\protect\citeauthoryear{??}{ela}{[n. d.]}]%
        {elasticsearch:url}
 \bibinfo{year}{[n. d.]}\natexlab{}.
\newblock \bibinfo{title}{Elasticsearch}.
\newblock \bibinfo{howpublished}{\url{https://www.elastic.co/}}.
  (\bibinfo{year}{[n. d.]}).
\newblock


\bibitem[\protect\citeauthoryear{??}{lat}{[n. d.]}]%
        {latency:url}
 \bibinfo{year}{[n. d.]}\natexlab{}.
\newblock \bibinfo{title}{Latency numbers every programmer should know}.
\newblock \bibinfo{howpublished}{\url{https://gist.github.com/jboner/2841832}}.
    (\bibinfo{year}{[n. d.]}).
\newblock


\bibitem[\protect\citeauthoryear{??}{luc}{[n. d.]b}]%
        {luceneutil:url}
 \bibinfo{year}{[n. d.]}\natexlab{b}.
\newblock \bibinfo{title}{luceneutil: lucene benchmarking utilities}.
\newblock
  \bibinfo{howpublished}{\url{https://github.com/mikemccand/luceneutil}}.
  (\bibinfo{year}{[n. d.]}).
\newblock


\bibitem[\protect\citeauthoryear{??}{pme}{[n. d.]}]%
        {pmem:url}
 \bibinfo{year}{[n. d.]}\natexlab{}.
\newblock \bibinfo{title}{Persistent Memory Wiki}.
\newblock \bibinfo{howpublished}{\url{https://nvdimm.wiki.kernel.org/start}}.
  (\bibinfo{year}{[n. d.]}).
\newblock


\bibitem[\protect\citeauthoryear{??}{wik}{[n. d.]}]%
        {wikipedia_data}
 \bibinfo{year}{[n. d.]}\natexlab{}.
\newblock \bibinfo{title}{Wikipedia:Database download}.
\newblock
  \bibinfo{howpublished}{\url{https://en.wikipedia.org/wiki/Wikipedia:Database_download}}.
    (\bibinfo{year}{[n. d.]}).
\newblock


\bibitem[\protect\citeauthoryear{Arulraj and Pavlo}{Arulraj and Pavlo}{2017}]%
        {arulraj17}
\bibfield{author}{\bibinfo{person}{Joy Arulraj} {and} \bibinfo{person}{Andrew
  Pavlo}.} \bibinfo{year}{2017}\natexlab{}.
\newblock \showarticletitle{How to Build a Non-Volatile Memory Database
  Management System}. In \bibinfo{booktitle}{{\em Proceedings of the 2017 ACM
  International Conference on Management of Data}} {\em
  (\bibinfo{series}{SIGMOD '17})}. \bibinfo{pages}{1753--1758}.
\newblock
\showURL{%
\url{http://db.cs.cmu.edu/papers/2017/p1753-arulraj.pdf}}


\bibitem[\protect\citeauthoryear{Białecki, Muir, and Ingersoll}{Białecki
  et~al\mbox{.}}{2012}]%
        {lucene4}
\bibfield{author}{\bibinfo{person}{Andrzej Białecki}, \bibinfo{person}{Robert
  Muir}, {and} \bibinfo{person}{Grant Ingersoll}.}
  \bibinfo{year}{2012}\natexlab{}.
\newblock \showarticletitle{Apache Lucene 4}. In \bibinfo{booktitle}{{\em
  Workshop on Open Source Information Retrieval}} {\em (\bibinfo{series}{SIGIR
  2012})}. \bibinfo{pages}{17--24}.
\newblock
\showISBNx{9780473220266}


\bibitem[\protect\citeauthoryear{DeBrabant, Arulraj, Pavlo, Stonebraker,
  Zdonik, and Dulloor}{DeBrabant et~al\mbox{.}}{2014}]%
        {debrabant14}
\bibfield{author}{\bibinfo{person}{Justin DeBrabant}, \bibinfo{person}{Joy
  Arulraj}, \bibinfo{person}{Andrew Pavlo}, \bibinfo{person}{Michael
  Stonebraker}, \bibinfo{person}{Stan Zdonik}, {and}
  \bibinfo{person}{Subramanya Dulloor}.} \bibinfo{year}{2014}\natexlab{}.
\newblock \showarticletitle{A Prolegomenon on OLTP Database Systems for
  Non-Volatile Memory}. In \bibinfo{booktitle}{{\em ADMS@VLDB}}.
  \bibinfo{pages}{57--63}.
\newblock
\showURL{%
\url{http://hstore.cs.brown.edu/papers/hstore-nvm.pdf}}


\bibitem[\protect\citeauthoryear{Islam, Wasi-ur Rahman, Lu, and Panda}{Islam
  et~al\mbox{.}}{2016}]%
        {islam:ics16}
\bibfield{author}{\bibinfo{person}{Nusrat~Sharmin Islam}, \bibinfo{person}{Md.
  Wasi-ur Rahman}, \bibinfo{person}{Xiaoyi Lu}, {and}
  \bibinfo{person}{Dhabaleswar~K. Panda}.} \bibinfo{year}{2016}\natexlab{}.
\newblock \showarticletitle{High Performance Design for HDFS with
  Byte-Addressability of NVM and RDMA}. In \bibinfo{booktitle}{{\em Proceedings
  of the 2016 International Conference on Supercomputing}} {\em
  (\bibinfo{series}{ICS '16})}. \bibinfo{publisher}{ACM}, \bibinfo{address}{New
  York, NY, USA}, Article \bibinfo{articleno}{8}, \bibinfo{numpages}{14}~pages.
\newblock
\showISBNx{978-1-4503-4361-9}
\showDOI{%
\url{https://doi.org/10.1145/2925426.2926290}}


\bibitem[\protect\citeauthoryear{McCandless, Hatcher, and
  Gospodnetic}{McCandless et~al\mbox{.}}{2010}]%
        {lucene_in_action:book}
\bibfield{author}{\bibinfo{person}{Michael McCandless}, \bibinfo{person}{Erik
  Hatcher}, {and} \bibinfo{person}{Otis Gospodnetic}.}
  \bibinfo{year}{2010}\natexlab{}.
\newblock \bibinfo{booktitle}{{\em Lucene in Action, Second Edition: Covers
  Apache Lucene 3.0}}.
\newblock \bibinfo{publisher}{Manning Publications Co.},
  \bibinfo{address}{Greenwich, CT, USA}.
\newblock
\showISBNx{1933988177, 9781933988177}


\bibitem[\protect\citeauthoryear{Nanavati, Schwarzkopf, Wires, and
  Warfield}{Nanavati et~al\mbox{.}}{2016}]%
        {acmq:nvm}
\bibfield{author}{\bibinfo{person}{Mihir Nanavati}, \bibinfo{person}{Malte
  Schwarzkopf}, \bibinfo{person}{Jake Wires}, {and} \bibinfo{person}{Andrew
  Warfield}.} \bibinfo{year}{2016}\natexlab{}.
\newblock \showarticletitle{Non-volatile Storage - Implications of the
  Datacenter's Shifting Center}. In \bibinfo{booktitle}{{\em ACM Queue, Volume
  13, Issue 9}}.
\newblock


\end{thebibliography}

\end{document}